\documentclass[conference]{IEEEtran}

\usepackage[letterpaper]{geometry}
\newgeometry{left=1.57cm,right=1.57cm,bottom=2.54cm,top=2.0cm} 
\IEEEoverridecommandlockouts

\usepackage{cite}
\usepackage{amsmath,amssymb,amsfonts,amsthm,enumitem}

\usepackage[ruled,linesnumbered]{algorithm2e}
\usepackage{algorithmic}
\usepackage{graphicx}
\usepackage{textcomp}
\usepackage{float}
\usepackage{placeins}
\usepackage{multirow}
\usepackage{dblfloatfix}
\usepackage{pifont}
\usepackage[table,xcdraw]{xcolor}
\usepackage[normalem]{ulem}
\useunder{\uline}{\ul}{}
\usepackage{booktabs}
\usepackage[none]{hyphenat}

\usepackage{adjustbox}
\usepackage{enumitem}
\usepackage[english]{babel}
\usepackage[utf8]{inputenc}
\usepackage[colorinlistoftodos]{todonotes}
\usepackage{balance}
\usepackage{comment}

\usepackage{pgfplots}
\usepackage{subcaption}

\begin{document}
	
\title{Is Approximation Universally Defensive Against Adversarial Attacks in Deep Neural Networks?}
\author{\IEEEauthorblockN{Ayesha Siddique, Khaza Anuarul Hoque}
\IEEEauthorblockA{\textit{Department of Electrical Engineering and Computer Science}\\ \textit{University of Missouri,
Columbia, MO, USA}\\
ayesha.siddique@mail.missouri.edu, hoquek@missouri.edu}
}

\maketitle
\begin{abstract}
Approximate computing is known for its effectiveness in improvising the energy efficiency of deep neural network (DNN) accelerators at the cost of slight accuracy loss. Very recently, the inexact nature of approximate components, such as approximate multipliers have also been reported successful in defending adversarial attacks on DNNs models. Since the approximation errors traverse through the DNN layers as masked or unmasked, this raises a key research question---\textit{can approximate computing always offer a defense against adversarial attacks in DNNs, i.e., are they universally defensive?} Towards this, we present an extensive adversarial robustness analysis of different approximate DNN accelerators (AxDNNs) using the state-of-the-art approximate multipliers. In particular, we evaluate the impact of ten adversarial attacks on different AxDNNs using the MNIST and CIFAR-10 datasets. Our results demonstrate that adversarial attacks on AxDNNs can cause 53\% accuracy loss whereas the same attack may lead to almost no accuracy loss (as low as 0.06\%) in the accurate DNN. Thus, approximate computing cannot be referred to as a universal defense strategy against adversarial attacks. 
\end{abstract}
	
\begin{IEEEkeywords}
		Adversarial Attacks, Adversarial Robustness, Approximate Computing, Deep Neural Networks
\end{IEEEkeywords}

\maketitle 

\section{Introduction}
\label{sec:introduction}

Approximate computing in deep neural networks (DNNs) has recently gained prominence in exploring the accuracy and energy trade-offs for big-data automation \cite{siddique2021exploring}. Approximate deep neural networks (AxDNN) accelerators employ inexact full adders \cite{riaz2020caxcnn}, truncated carry chains \cite{hanif2019cann}, etc., which induce approximation errors in them. Unfortunately, DNNs are susceptible to adversarial attacks~\cite{xu2021security} and AxDNNs are no exception \cite{guesmi2020defensive}. This limits their deployment in the safety-critical applications since the adversary may use partial information about the model to craft adversarial examples and exploit transferability property of DNNs \cite{papernot2017practical} \cite{demontis2019adversarial} to attack AxDNNs. Recent works in defending adversarial attacks are targeted mostly for accurate DNNs \cite{wang2020dnnguard} \cite{xu2021security} and thus, the robustness and defense of AxDNNs is vastly under-explored \cite{capra2020updated}. 

Very recently, Guesmi et al. presented approximate computing as an effective structural defense strategy against adversarial attacks by incorporating an array multiplier, with approximate mirror adders instead of exact full adders, in the AxDNN inference phase \cite{guesmi2020defensive}. Even though this solution opens up a new dimension of research, such defensive behavior of approximate computing cannot be generalized with one AxDNN. This is due to the fact that approximation errors traverse through the AxDNN layers as masked and un-masked and hence, may render their structural defense inconsistent in an adversarial environment. Hence, there is a pent-up need to explore the adversarial robustness of AxDNNs extensively which also includes investigating the impact of adversarial attacks with different perturbation budgets under different attack scenarios. Additionally, it is also required to explore if quantization is supportive towards adversarial defense in AxDNNs since quantization can also improve the robustness in accurate DNNs~\cite{khalid2019qusecnets}. Towards this, the research questions that need to be investigated are as follows: \\

 \noindent \textbf{(Q1)} Does approximate computing in AxDNNs provide universal defense against adversarial attacks? How does the adversarial robustness of AxDNNs vary with the change in the perturbation budget?\\
 \noindent \textbf{(Q2)} Are adversarial attacks transferable from accurate DNNs to AxDNNs irrespective of their difference in exactness and model structure?\\
\noindent \textbf{(Q3)} How does approximate computing react to quantization in AxDNNs under adversarial attacks? Are they supportive or antagonistic to each other?

\begin{figure}[!t]
	\centering
	\vspace{-0.01in}
	\includegraphics[width=0.95\linewidth]{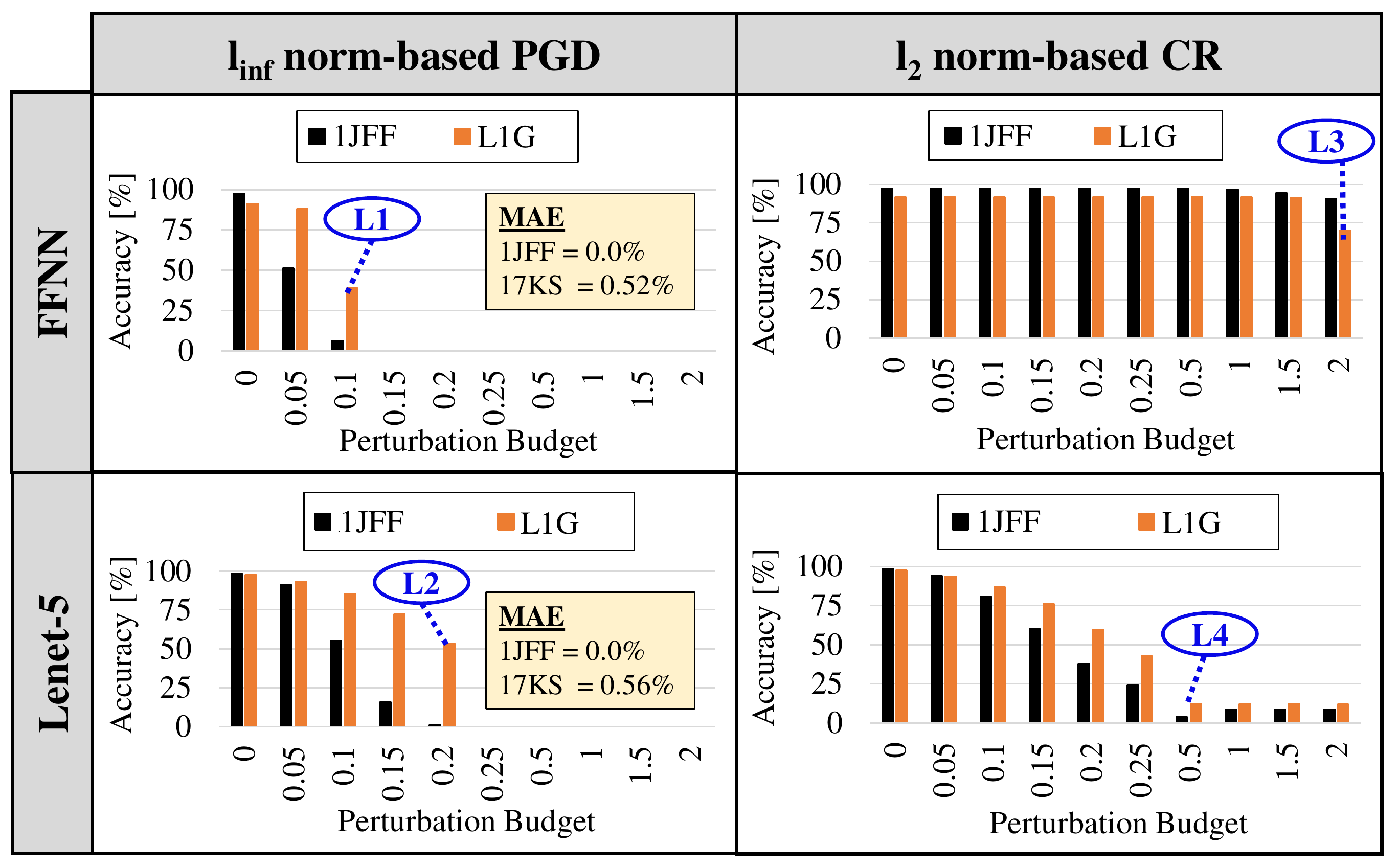}
	\vspace{-0.04in}
	\caption{Impact of adversarial attacks on accurate and approximate versions of FFNN and Lenet-5. The accurate and approximate DNNs contain accurate 1JFF and approximate L1G multipliers, respectively, from Evoapprox8b \cite{mrazek2017evoapproxsb} library.}
	\label{fig:mlplenet5}
	\vspace{-0.17in}
\end{figure}

\subsection{Motivational Case Study and Key Observations}
Prior to extensively analyzing the adversarial robustness of AxDNNs, we first present a motivational case study to demonstrate their both defensive and perturbing nature towards adversarial attacks. We trained a 5-layered convolutional neural network, i.e., Lenet-5, and feed-forward neural network (FFNN) on the MNIST \cite{cohen2017emnist} dataset. We replaced their accurate multipliers with approximate counterparts, using Evoapprox8b \cite{mrazek2017evoapproxsb} library, to build two different AxDNNs. We compared the classification accuracy of each resulting AxDNN with its exact counterpart under the $\mathrm{\textit{l}}_\mathrm{\infty}$ norm-based projected gradient descent (PGD) and $\mathrm{\textit{l}}_\mathrm{2}$ norm-based contrast reduction (CR) attacks. As shown in Fig. \ref{fig:mlplenet5}, we observe that the accuracy of both AxDNNs is higher than the accurate DNNs in the case of former attack (see label L1 and L2). However, the same approximate FFNN exhibits an opposite behavior in the case of later attack i.e., its accuracy decreases with an increase in the strength of the attack (see label L3). Moreover, a clear drop in accuracy of more than 75\% is observed around perturbation budget ($\epsilon$) of 0.5 in the case of approximate Lenet-5 (AxL5) with the later attack (see Label L4). After this value, the accuracy of AxL5 decreases sharply. Such conflicting observations motivated us to extensively analyze the adversarial robustness of AxDNNs.

\subsection{Novel Contributions}
This paper makes the following novel contributions: 

\begin{enumerate}
\item An extensive adversarial approximation analysis to expose the perturbing nature of approximation noise in different adversarial settings with varying perturbation budgets. \textbf{[Section IV.B]}
\item A transferability analysis to determine whether the adversarial attacks are transferable from accurate DNNs to AxDNNs irrespective of their difference in exactness and model structure. \textbf{[Section IV.C]}
\item An adversarial quantization analysis to determine whether quantization and approximate computing are supportive or antagonistic to each other. \textbf{[Section IV.D]} 

\end{enumerate} 

\vspace{0.1in}
Since the multipliers consume more energy as compared to other arithmetic units (e.g., adders) \cite{marchisio2020red}; therefore, we employ the state-of-the-art approximate multipliers \cite{mrazek2017evoapproxsb} in AxDNNs. In particular, we explore the impact of 10 different adversarial attacks on approximate Lenet-5 (AxL5) and Alexnet (AxAlx). We use the MNIST \cite{cohen2017emnist} and CIFAR-10 \cite{krizhevsky2009learning} datasets for the adversarial robustness analysis. Our results demonstrate that an adversarial attack on AxDNNs may lead to 53\% accuracy loss. Conversely, the same attack may lead to almost no accuracy loss (as low as 0.06\%) in accurate DNNs. This behavior contradicts the observations in \cite{guesmi2020defensive}. Our analysis reveals that \textit{AxDNNs are not universally defensive towards the adversarial attacks.} Furthermore, the adversarial attacks are transferable from accurate DNNs to AxDNNs irrespective of their difference in exactness and model structure. \textit{We also observe that approximate computing acts antagonistically to quantization}. 

\vspace{0.05in}
The remainder of this paper is structured as follows: Section \ref{sec:threatModel} and Section \ref{sec:methodology} present a threat model and methodology for analyzing the adversarial robustness of AxDNNs. Section \ref{sec:results} presents the results for adversarial robustness analysis of AxDNNs in comparison with accurate DNNs. Finally, Section \ref{sec:conclusion} concludes the paper.

\vspace{0.1in}
\section{Threat Model}
\label{sec:threatModel}
In this section, a threat model is presented for exploring the adversarial robustness of AxDNNs.

\subsection{Adversary's Knowledge}
\label{subsec:advknow}
\begin{figure}[!h]
	\centering
	\vspace{-0.01in}
	\includegraphics[width=1\linewidth]{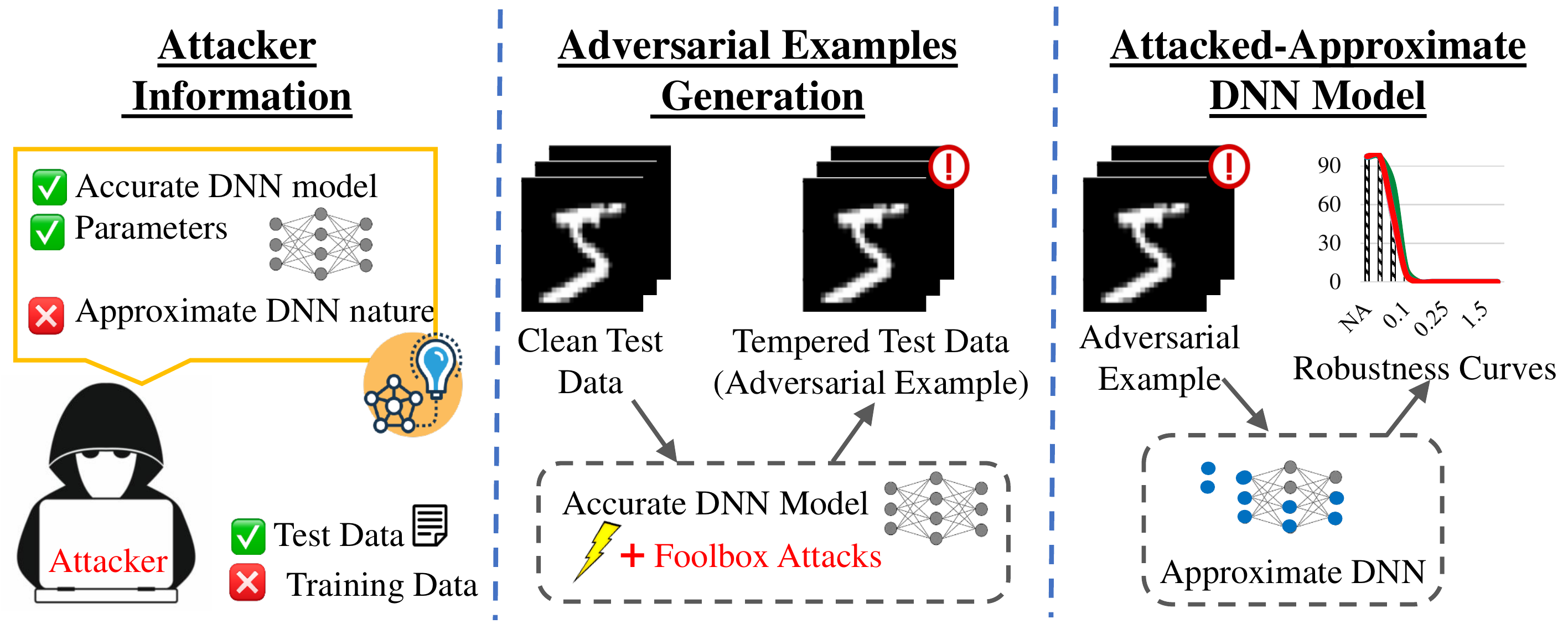}
	\vspace{-0.2in}
	\caption{Attack scenario with both model structure and inexactness not known to the adversary}
	\label{fig:attackScenario}
	\vspace{-0.1in}
\end{figure}

We assume that the adversary uses an accurate classifier model for generating the adversarial examples. The adversary has either (i) partial knowledge about the AxDNN i.e., the model structure is known but inexactness is not known, or (ii) no knowledge about the AxDNN i.e., both model structure and inexactness are not known (see Fig. \ref{fig:attackScenario}). Since the adversary lacks the information about the inexactness of AxDNNs only in the former case; therefore, it is considered as a special case of transferability. This attack scenario is similar to the black-box attacks in \cite{guesmi2020defensive}. 

\begin{table}[!b]
\centering
\caption {Adversarial Attacks \cite{rauber2017foolbox}, Types, and Distance Metrics} 
\label{tab:attacks}
\begin{tabular}{|l|l|l|l|}
\hline & \\[-2.5ex]
\rowcolor[HTML]{E0E0E0} 
Attack Name  & Attack  & Distance                              \\
\rowcolor[HTML]{E0E0E0} 
             & Type            & Measure                               \\\hline&\\[-2.5ex]
Fast Gradient Method (FGM)           & gradient        & $\mathrm{\textit{l}}_\mathrm{2}$, $\mathrm{\textit{l}}_\mathrm{\infty}$ norm        \\\hline&\\[-2.5ex]
Basic Iterative Method (BIM)         & gradient        & $\mathrm{\textit{l}}_\mathrm{2}$, $\mathrm{\textit{l}}_\mathrm{\infty}$  norm                \\\hline&\\[-2.5ex]
Projected Gradient Descent (PGD)         & gradient        & $\mathrm{\textit{l}}_\mathrm{2}$, $\mathrm{\textit{l}}_\mathrm{\infty}$  norm          \\\hline&\\[-2.5ex]
Contrast Reduction Attack (CR)          & decision        & $\mathrm{\textit{l}}_\mathrm{2}$ norm      \\\hline&\\[-2.5ex]
Repeated Additive Gaussian (RAG)          & decision        & $\mathrm{\textit{l}}_\mathrm{2}$ norm      \\\hline&\\[-2.5ex]
Repeated Additive Uniform Noise (RAU)          & decision        & $\mathrm{\textit{l}}_\mathrm{2}$ , $\mathrm{\textit{l}}_\mathrm{\infty}$  norm       \\\hline
\end{tabular}
\vspace{-0.01in}
\end{table}

\subsection{Attack Generation}
In this paper, the adversary is assumed to be exploratory. The adversary can evade the AxDNN by tampering with the test images, during the inference phase, without influencing the training data. The adversary seeks to craft adversarial examples by finding the perturbation that maximizes the loss of a model on a given sample while keeping the perturbation magnitude lower than a given budget \cite{khalid2019trisec}. Table \ref{tab:attacks} enlists the gradient and decision-based attacks and the distance metrics used in this paper. The distance metrics such as, $l_0$, $l_2$, and $l_\infty$ norms help in approximating the human perception of visual difference. The $l_0$ norm counts the number of pixels with different values at corresponding positions in the original and perturbed images. The $l_2$ norm measures the Euclidean distance between two images. The $l_\infty$ norm measures the maximum difference for all pixels at corresponding positions in two images.

\section{Evaluation Methodology}
\label{sec:methodology}

\begin{figure}[!t]
	\centering
	\vspace{-0.01in}
	\includegraphics[width=1\linewidth]{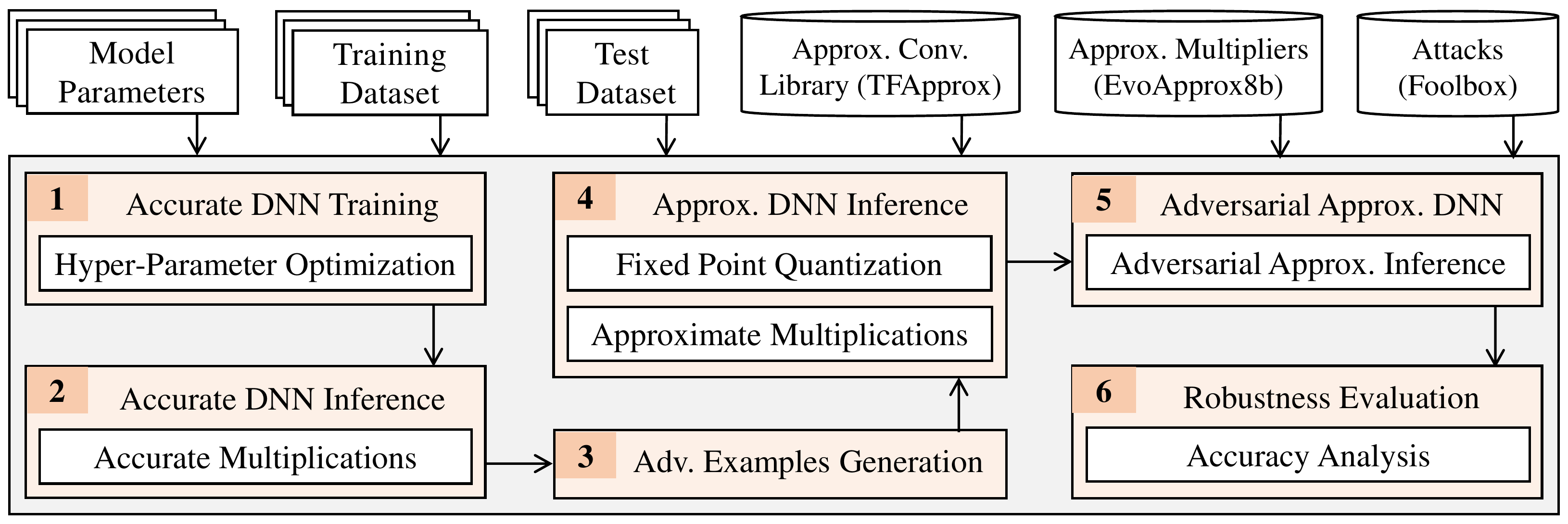}
	\vspace{-0.1in}
	\caption{Methodology for analyzing the adversarial robustness of approximate deep neural networks (AxDNNs)}
	\label{fig:methodology}
	\vspace{-0.2in}
\end{figure}

Fig. \ref{fig:methodology} shows the overview of our methodology for AxDNN's robustness evaluation. It consists of four main steps: accurate DNN training, adversarial examples generation, attacks on AxDNN inference, and percentage robustness. Algorithm \ref{alg:robust} delineates these steps. Line 1 and 2 train the accurate DNN with accurate multipliers and check whether the accuracy of the trained model is above the user-defined threshold. In this paper, we consider baseline accuracy as a threshold value. The adversarial robustness analysis of accurate DNN and AxDNNs starts from Line 6. First, the accurate multiplier and different adversarial attacks, with multiple perturbation budgets (ranging from 0 to \textit{p}, where \textit{p} is a set of integers) are used for generating the adversarial examples. Higher is the perturbation budget, the higher is the strength of the adversarial attack. Then, the quantized accurate DNN and AxDNNs with accurate and approximate multipliers, respectively, are evaluated against the adversarial examples. Line 8 verifies if the adversary succeeded in misclassification i.e., forcing the output to an arbitrary false label. If the goal of the adversary is achieved then, the counter of successful attack generation is incremented. Lastly, the robustness is evaluated in Line 15, for every perturbation budget, as the percentage rate of attacks for which the adversary fails to generate an effective adversarial example that fools the victim accurate DNN or AxDNN. 

\SetInd{0.5em}{0.5em}
\begin{algorithm}[!h]
	\caption{Robustness Evaluation}
	\label{alg:robust}
        \small
		\footnotesize
        \DontPrintSemicolon

        \SetKwInOut{Input}{Inputs}\SetKwInOut{Output}{Outputs}
        \Input{Type of multipliers: mults =\{ACC, JV3,  ...\}; \\ 
        Type of adversarial attack: attack = BIM or PGD, etc. \\
        Perturbation budget: eps = [0, p]; \\
        Labelled test set: $\mathcal{D}$ = ($X^t$, $L^t$); \\
        Quantization level: Qlevel; \\
        Accuracy threshold: Ath}

		\Output{Percentage Robustness: Rlevels}

		\begin{algorithmic}[1]
        \STATE model = ExactDNNtrain (mults(1)) \\
        // Train DNN with accurate multiplier 
        \IF {Accuracy(model) $\ge$ Ath}
            \FOR {j = 1 : length(eps)}{
                \STATE adv = 0;
                \FOR {k = 1 : size($\mathcal{D}$)}{
                    \STATE ($X^{t*}_k$, $L^{t*}_k$) = AdvExGen (model, mults(1), eps(j), attack, $X^t_k$) \\ 
                    // Adv. examples generation with accurate multiplier
                    \STATE Qmdl = FixedPointQuantization (model, Qlevel) \\
                    // Apply fixed point quantization on inference model
                   \STATE ($X^{q}_k$, $L^{q}_k$) = AdvAttackOnQuanModel (Qmdl, mults eps(j), attack, $X^{t*}_k$, $L^{t*}_k$) \\ 
                    // Adv. attacks on accurate DNN and AxDNNs 
                    \IF {$L^{q}_k$ $\neq$ $L^t_k$}{
                        \STATE adv++;}
                    \ELSE{
                        \STATE NOP;}
                    \ENDIF}
                \ENDFOR    
                \STATE Rlevels (eps}(j)) = (1 - Adv/ size($\mathcal{D}$)) * 100;
            \ENDFOR
        \ENDIF

\end{algorithmic}
\end{algorithm}

\begin{figure*}[!t]
\vspace{-0.01in}
\centering
	\begin{subfigure}[!b]{.28\textwidth}
		\centering
		\vspace{-0.01in}
		\caption{$\mathrm{\textit{l}}_\mathrm{\infty}$-based BIM Attack}
		{\resizebox{\textwidth}{!}{\definecolor{mycolor2}{rgb}{0.00000,0.44700,0.74100}%
\definecolor{mycolor1}{rgb}{0.85000,0.32500,0.09800}%
\definecolor{mycolor3}{rgb}{0,128,0}%

\begin{tikzpicture}[scale=0.6]

\node at (5,-12) {Multipliers};
\node [rotate=90] at (-1.3,-5) {Perturbation Budget ($\epsilon$)};


\node at (1,-11) {M1};
\node at (2,-11) {M2};
\node at (3,-11) {M3};
\node at (4,-11) {M4};
\node at (5,-11) {M5};
\node at (6,-11) {M6};
\node at (7,-11) {M7};
\node at (8,-11) {M8};
\node at (9,-11) {M9};

\node at (0,-1) {0.00 -};
\node at (0,-2) {0.05 -};
\node at (0,-3) {0.10 -};
\node at (0,-4) {0.15 -};
\node at (0,-5) {0.20 -};
\node at (0,-6) {0.25 -};
\node at (0,-7) {0.50 -};
\node at (0,-8) {1.00 -};
\node at (0,-9) {1.50 -};
\node at (0,-10) {2.00 -};

\foreach \y [count=\n] in {
      {98,98,98,96,96,91,96,90,93},
      {97,96,96,93,94,73,92,84,74},
      {93,90,90,85,85,70,83,71,72},
      {77,72,77,71,75,67,63,45,77},
      {54,50,56,51,56,49,40,23,25},
      {0,0,0,0,0,0,0,0,0},
      {0,0,0,0,0,0,0,0,0},
      {0,0,0,0,0,0,0,0,0},
      {0,0,0,0,0,0,0,0,0},
      {0,0,0,0,0,0,0,0,0},
    } {
      \foreach \x [count=\m] in \y {
        \node[fill=violet!\x!cyan!\x!cyan!\x!cyan!\x!cyan!\x!green!\x!green!\x!black!\x!gray, minimum size=6mm, text=white] at (\m,-\n) {\x};     
      }
    }
    
\end{tikzpicture}}\label{fig:linfBIM}}
	\end{subfigure}
	\begin{subfigure}[!b]{.23\textwidth}
		\centering
		 \vspace{-0.01in}
		\caption{$\mathrm{\textit{l}}_\mathrm{2}$-based BIM Attack}
		{\resizebox{\textwidth}{!}{\definecolor{mycolor2}{rgb}{0.00000,0.44700,0.74100}%
\definecolor{mycolor1}{rgb}{0.85000,0.32500,0.09800}%
\definecolor{mycolor3}{rgb}{0,128,0}%

\begin{tikzpicture}[scale=0.6]

\node at (5,-12) {Multipliers};

\node at (1,-11) {M1};
\node at (2,-11) {M2};
\node at (3,-11) {M3};
\node at (4,-11) {M4};
\node at (5,-11) {M5};
\node at (6,-11) {M6};
\node at (7,-11) {M7};
\node at (8,-11) {M8};
\node at (9,-11) {M9};

\foreach \y [count=\n] in {
      {98,98,98,96,96,91,96,90,93},
      {98,98,98,96,96,91,96,90,93},
      {98,98,98,96,97,91,95,90,93},
      {98,98,98,96,96,91,95,90,92},
      {98,98,98,96,96,90,95,90,91},
      {98,97,97,96,96,90,95,89,89},
      {97,96,97,94,95,88,93,87,84},
      {94,92,93,88,90,80,86,77,75},
      {86,82,83,77,81,70,75,64,64},
      {69,65,68,62,66,57,58,48,49},
    } {
      \foreach \x [count=\m] in \y {
        \node[fill=violet!\x!cyan!\x!cyan!\x!cyan!\x!cyan!\x!green!\x!green!\x!black!\x!gray, minimum size=6mm, text=white] at (\m,-\n) {\x};               
      }
    }
\end{tikzpicture}}\label{fig:l2BIM}}
	\end{subfigure}
	\begin{subfigure}[!b]{.23\textwidth}
		\centering
		 \vspace{-0.01in}
		\caption{$\mathrm{\textit{l}}_\mathrm{\infty}$-based FGM Attack}		{\resizebox{\textwidth}{!}{\definecolor{mycolor2}{rgb}{0.00000,0.44700,0.74100}%
\definecolor{mycolor1}{rgb}{0.85000,0.32500,0.09800}%
\definecolor{mycolor3}{rgb}{0,128,0}%

\begin{tikzpicture}[scale=0.6]

\node at (5,-12) {Multipliers};

\node at (1,-11) {M1};
\node at (2,-11) {M2};
\node at (3,-11) {M3};
\node at (4,-11) {M4};
\node at (5,-11) {M5};
\node at (6,-11) {M6};
\node at (7,-11) {M7};
\node at (8,-11) {M8};
\node at (9,-11) {M9};

\foreach \y [count=\n] in {
      {98,98,98,96,96,91,96,90,93},
      {97,97,96,94,94,87,93,86,71},
      {94,93,93,87,87,73,88,79,77},
      {89,86,86,76,79,70,78,65,83},
      {77,75,73,60,68,53,65,52,41},
      {61,59,57,42,49,34,59,41,53},
      {11,12,12,12,12,12,10,12,10},
      {10,10,11,12,12,12,9,11,9},
      {10,10,11,12,12,12,9,11,9},
      {10,10,11,12,12,12,9,11,9},
    } {
      \foreach \x [count=\m] in \y {
        \node[fill=violet!\x!cyan!\x!cyan!\x!cyan!\x!cyan!\x!green!\x!green!\x!black!\x!gray, minimum size=6mm, text=white] at (\m,-\n) {\x};         
      }
    }
\end{tikzpicture}}\label{fig:linfFGM}}
	\end{subfigure}
	\begin{subfigure}[!b]{.23\textwidth}
		\centering
	    \vspace{-0.01in}
		\caption{$\mathrm{\textit{l}}_\mathrm{2}$-based FGM Attack}
		{\resizebox{\textwidth}{!}{\definecolor{mycolor2}{rgb}{0.00000,0.44700,0.74100}%
\definecolor{mycolor1}{rgb}{0.85000,0.32500,0.09800}%
\definecolor{mycolor3}{rgb}{0,128,0}%

\begin{tikzpicture}[scale=0.6]

\node at (5,-12) {Multipliers};

\node at (1,-11) {M1};
\node at (2,-11) {M2};
\node at (3,-11) {M3};
\node at (4,-11) {M4};
\node at (5,-11) {M5};
\node at (6,-11) {M6};
\node at (7,-11) {M7};
\node at (8,-11) {M8};
\node at (9,-11) {M9};

\foreach \y [count=\n] in {
      {98,98,98,96,96,91,96,90,93},
      {98,98,98,96,96,91,96,90,93},
      {98,98,98,96,96,91,95,90,93},
      {98,98,98,96,96,91,95,90,93},
      {98,98,98,96,96,90,95,90,98},
      {98,98,98,96,96,90,95,89,98},
      {98,97,97,95,96,89,94,88,97},
      {96,95,95,92,83,84,97,83,81},
      {94,92,92,87,89,78,86,76,73},
      {89,97,87,79,82,71,80,70,65},
    } {
      \foreach \x [count=\m] in \y {
        \node[fill=violet!\x!cyan!\x!cyan!\x!cyan!\x!cyan!\x!green!\x!green!\x!black!\x!gray, minimum size=6mm, text=white] at (\m,-\n) {\x};        
      }
    }
\end{tikzpicture}}\label{fig:l2FGM}}
	\end{subfigure}
    \vspace{-0.07in}
	\caption{Adversarial robustness of accurate and approximate LeNet-5 under BIM and FGM attacks with the MNIST \cite{cohen2017emnist} dataset. The labels M1 to M9 refer to the 1JFF (Accurate), 96D, 12N4, 17KS, 1AGV, FTA, JQQ, L40 and JV3 multipliers in EvoApprox8b \cite{mrazek2017evoapproxsb} library.} 
	\label{fig:LenetBIMFGM}
	\vspace{-0.1in}
\end{figure*}
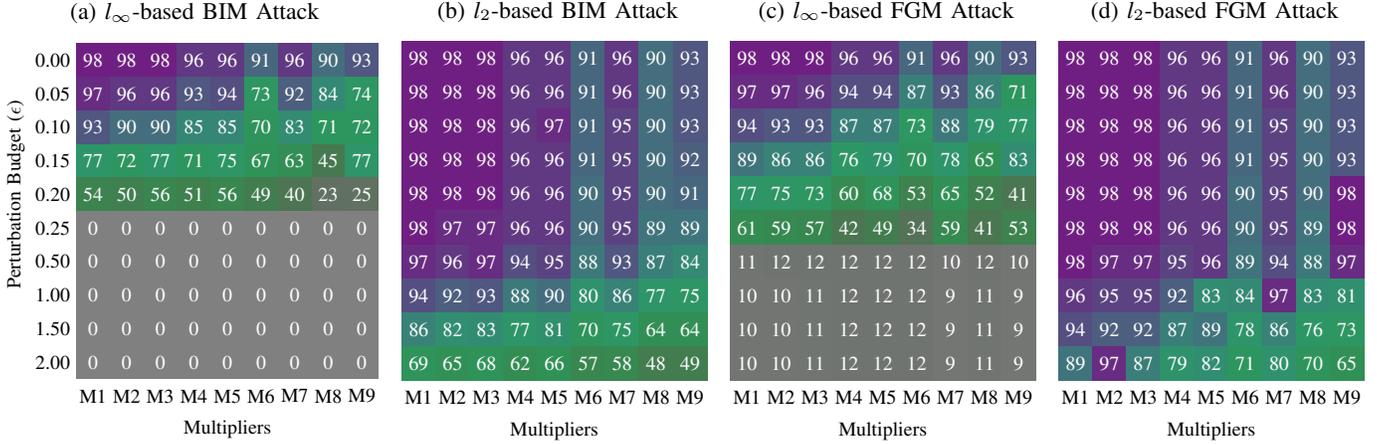

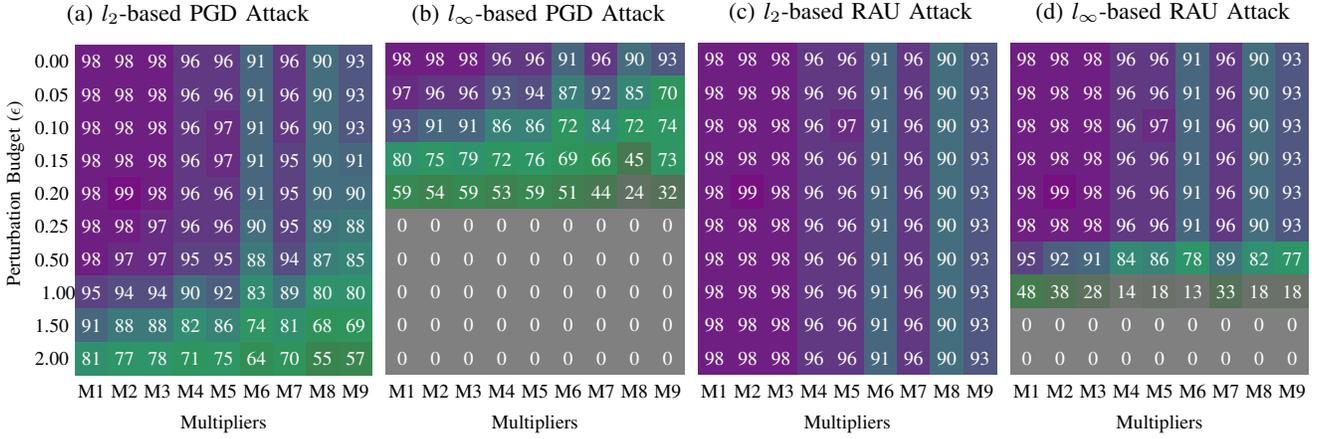
\begin{figure*}[!t]
\centering
	\begin{subfigure}[b]{.28\textwidth}
		\centering
		\vspace{-0.01in}		\caption{$\mathrm{\textit{l}}_\mathrm{2}$-based PGD Attack}
		{\resizebox{\textwidth}{!}{\def\colorModel{hsb} \definecolor{mycolor2}{hsb}{0.00000,0.44700,0.74100}%
\definecolor{mycolor1}{hsb}{0.85000,0.32500,0.09800}%
\definecolor{mycolor3}{hsb}{0,128,0}%

\begin{tikzpicture}[scale=0.6]

\node at (5,-12) {Multipliers};
\node [rotate=90] at (-1.3,-5) {Perturbation Budget ($\epsilon$)};

\node at (1,-11) {M1};
\node at (2,-11) {M2};
\node at (3,-11) {M3};
\node at (4,-11) {M4};
\node at (5,-11) {M5};
\node at (6,-11) {M6};
\node at (7,-11) {M7};
\node at (8,-11) {M8};
\node at (9,-11) {M9};

\node at (0,-1) {0.00 -};
\node at (0,-2) {0.05 -};
\node at (0,-3) {0.10 -};
\node at (0,-4) {0.15 -};
\node at (0,-5) {0.20 -};
\node at (0,-6) {0.25 -};
\node at (0,-7) {0.50 -};
\node at (0,-8) {1.00 };
\node at (0,-9) {1.50 -};
\node at (0,-10) {2.00 -};

\foreach \y [count=\n] in {
      {98,98,98,96,96,91,96,90,93},
      {98,98,98,96,96,91,96,90,93},
      {98,98,98,96,97,91,96,90,93},
      {98,98,98,96,97,91,95,90,91},
      {98,99,98,96,96,91,95,90,90},
      {98,98,97,96,96,90,95,89,88},
      {98,97,97,95,95,88,94,87,85},
      {95,94,94,90,92,83,89,80,80},
      {91,88,88,82,86,74,81,68,69},
      {81,77,78,71,75,64,70,55,57},
    } {
      \foreach \x [count=\m] in \y {
        \node[fill=violet!\x!cyan!\x!cyan!\x!cyan!\x!cyan!\x!green!\x!green!\x!black!\x!gray, minimum size=6mm, text=white] at (\m,-\n) {\x};            
      }
    }
\end{tikzpicture}}\label{fig:l2PGD}}
	\end{subfigure}%
	\begin{subfigure}[b]{.225\textwidth}
		\centering
		\caption{$\mathrm{\textit{l}}_\mathrm{\infty}$-based PGD Attack}
        \vspace{-0.01in}
		{\resizebox{\textwidth}{!}{\definecolor{mycolor2}{rgb}{0.00000,0.44700,0.74100}%
\definecolor{mycolor1}{rgb}{0.85000,0.32500,0.09800}%
\definecolor{mycolor3}{rgb}{0,128,0}%

\begin{tikzpicture}[scale=0.6]

\node at (5,-12) {Multipliers};

\node at (1,-11) {M1};
\node at (2,-11) {M2};
\node at (3,-11) {M3};
\node at (4,-11) {M4};
\node at (5,-11) {M5};
\node at (6,-11) {M6};
\node at (7,-11) {M7};
\node at (8,-11) {M8};
\node at (9,-11) {M9};

\foreach \y [count=\n] in {
      {98,98,98,96,96,91,96,90,93},
      {97,96,96,93,94,87,92,85,70},
      {93,91,91,86,86,72,84,72,74},
      {80,75,79,72,76,69,66,45,73},
      {59,54,59,53,59,51,44,24,32},
      {0,0,0,0,0,0,0,0,0},
      {0,0,0,0,0,0,0,0,0},
      {0,0,0,0,0,0,0,0,0},
      {0,0,0,0,0,0,0,0,0},
      {0,0,0,0,0,0,0,0,0},
    } {
      \foreach \x [count=\m] in \y {
        \node[fill=violet!\x!cyan!\x!cyan!\x!cyan!\x!cyan!\x!green!\x!green!\x!black!\x!gray, minimum size=6mm, text=white] at (\m,-\n) {\x};   
      }
    }
\end{tikzpicture}}\label{fig:linfPGD}}
	\end{subfigure}%
	\begin{subfigure}[b]{.225\textwidth}
		\centering
		\vspace{-0.01in}
		\caption{$\mathrm{\textit{l}}_\mathrm{2}$-based RAU Attack}
		{\resizebox{\textwidth}{!}{\definecolor{mycolor2}{rgb}{0.00000,0.44700,0.74100}%
\definecolor{mycolor1}{rgb}{0.85000,0.32500,0.09800}%
\definecolor{mycolor3}{rgb}{0,128,0}%

\begin{tikzpicture}[scale=0.6]

\node at (5,-12) {Multipliers};

\node at (1,-11) {M1};
\node at (2,-11) {M2};
\node at (3,-11) {M3};
\node at (4,-11) {M4};
\node at (5,-11) {M5};
\node at (6,-11) {M6};
\node at (7,-11) {M7};
\node at (8,-11) {M8};
\node at (9,-11) {M9};

\foreach \y [count=\n] in {
      {98,98,98,96,96,91,96,90,93},
      {98,98,98,96,96,91,96,90,93},
      {98,98,98,96,97,91,96,90,93},
      {98,98,98,96,96,91,96,90,93},
      {98,99,98,96,96,91,96,90,93},
      {98,98,98,96,96,91,96,90,93},
      {98,98,98,96,96,91,96,90,93},
      {98,98,98,96,96,91,96,90,93},
      {98,98,98,96,96,91,96,90,93},
      {98,98,98,96,96,91,96,90,93},
    } {
      \foreach \x [count=\m] in \y {
        \node[fill=violet!\x!cyan!\x!cyan!\x!cyan!\x!cyan!\x!green!\x!green!\x!black!\x!gray, minimum size=6mm, text=white] at (\m,-\n) {\x};           
      }
    }
\end{tikzpicture}}\label{fig:l2RAU}}
	\end{subfigure}
	\begin{subfigure}[b]{.225\textwidth}
		\centering
		\vspace{-0.01in}		\caption{$\mathrm{\textit{l}}_\mathrm{\infty}$-based RAU Attack}
		{\resizebox{\textwidth}{!}{\definecolor{mycolor2}{rgb}{0.00000,0.44700,0.74100}%
\definecolor{mycolor1}{rgb}{0.85000,0.32500,0.09800}%
\definecolor{mycolor3}{rgb}{0,128,0}%

\begin{tikzpicture}[scale=0.6]

\node at (5,-12) {Multipliers};

\node at (1,-11) {M1};
\node at (2,-11) {M2};
\node at (3,-11) {M3};
\node at (4,-11) {M4};
\node at (5,-11) {M5};
\node at (6,-11) {M6};
\node at (7,-11) {M7};
\node at (8,-11) {M8};
\node at (9,-11) {M9};

\foreach \y [count=\n] in {
      {98,98,98,96,96,91,96,90,93},
      {98,98,98,96,96,91,96,90,93},
      {98,98,98,96,97,91,96,90,93},
      {98,98,98,96,96,91,96,90,93},
      {98,99,98,96,96,91,96,90,93},
      {98,98,98,96,96,91,96,90,93},
      {95,92,91,84,86,78,89,82,77},
      {48,38,28,14,18,13,33,18,18},
      {0,0,0,0,0,0,0,0,0},
      {0,0,0,0,0,0,0,0,0},
    } {
      \foreach \x [count=\m] in \y {
        \node[fill=violet!\x!cyan!\x!cyan!\x!cyan!\x!cyan!\x!green!\x!green!\x!black!\x!gray, minimum size=6mm, text=white] at (\m,-\n) {\x};             
      }
    }
\end{tikzpicture}}\label{fig:linfRAU}}
	\end{subfigure}
	\vspace{-0.07in}	
	\caption{Adversarial robustness of accurate and approximate LeNet-5 under PGD and RAU attacks with the MNIST \cite{cohen2017emnist} dataset. The labels M1 to M9 refer to the 1JFF (Accurate), 96D, 12N4, 17KS, 1AGV, FTA, JQQ, L40 and JV3 multipliers in EvoApprox8b \cite{mrazek2017evoapproxsb} library.} 
	\label{fig:LenetPGDRAU}
	\vspace{-0.2in}
\end{figure*}

\section{Results and Discussions}
\label{sec:results}
This section discusses the experimental setup and the impact of adversarial attacks and quantization on AxDNNs under different adversarial settings (adversarial approximation and adversarial quantization analysis) and transferability of adversarial attacks from accurate DNNs to AxDNNs.

\subsection{Experimental Setup}
In this paper, accurate Lenet-5 (AccL5) and Alexnet (AccAlx) architectures are used with their baseline accuracy as 98\% and 81\%, respectively. The LeNet-5 architecture is comprised of two sets of convolutional and average pooling layers, followed by a flattening convolutional layer, two fully-connected layers, and finally a softmax classifier. The Alexnet architecture contains five convolutional layers, three average pooling layers, and two fully connected layers. For approximate counterparts of these accurate DNNs, the accurate multipliers in the convolutional layers are replaced with approximate unsigned multipliers using Evoapprox8b \cite{mrazek2017evoapproxsb} library. The approximate multipliers are employed in AxL5 and AxAlx according to their error resilience towards the MNIST \cite{cohen2017emnist} and CIFAR-10 \cite{krizhevsky2009learning} classification, respectively. For example, the approximate multipliers having accuracy less than 90\% in AxL5 and 75\% in AxAlx are discarded. The adversarial examples are generated using the Foolbox library \cite{rauber2017foolbox}.

\subsection{Adversarial Approximation Analysis}
\label{subsec:advapprox}
To investigate the adversarial robustness of AxDNNs, this section discusses the impact of both gradient and decision-based attacks with reference to the first attack scenario in Section \ref{subsec:advknow}.

\vspace{0.08in}
\subsubsection{\uline{Approximate DNNs under Gradient-Based Attacks}}

In AxDNNs, both approximation noise and adversarial robustness can be quantified in terms of the mean average error (MAE) of the approximate multipliers. The lower the MAE is, the higher is the actual inference accuracy of AxDNNs (in the absence of adversarial attacks) and hence, higher is their adversarial robustness. For example, Fig. \ref{fig:LenetBIMFGM} shows that M8-based (MAE = 1.54\%) AxL5 exhibits 6\% inference accuracy lower than M7-based (MAE = 1.12\%) AxL5 in absence of any adversarial attack. Hence, it undergoes 11\% more accuracy loss under $\mathrm{\textit{l}}_\mathrm{\infty}$ norm-based BIM attack with $\epsilon$ = 0.2. It is also observed that two AxDNNs, having the same inference accuracy, behave in a similar fashion under adversarial attacks. For instance, M2 and M3-based AxL5 perform close to AccL5 under adversarial attacks due to their same inference accuracy. 

The adversarial robustness analysis in Fig. \ref{fig:LenetBIMFGM} reveals that the accuracy and hence, adversarial robustness of AxDNNs decreases with an increase in the strength of the adversarial attacks, even if the adversary is unaware of the inexact inference engine. For example, $\mathrm{\textit{l}}_\mathrm{\infty}$ norm-based BIM attack, with $\epsilon$ = 0.2, on AccL5 leads to 44\% accuracy loss. However, the same strength of the attack results in 67\% accuracy loss in M8-based AxL5 (see Fig. 4a). The same trend is observed with other gradient-based attacks. 

It is also observed that AxDNNs, similar to accurate DNNs, exhibit more adversarial robustness under $\mathrm{\textit{l}}_\mathrm{2}$ norm-based attacks as compared to their $\mathrm{\textit{l}}_\mathrm{\infty}$-norm based counterparts. This trend is noticeable with all adversarial attacks. For example, $\mathrm{\textit{l}}_\mathrm{\infty}$ norm-based FGM attack, with $\epsilon$ = 0.25, leads to 37\% and 49\% accuracy loss in AccL5 and M8-based AxL5, respectively (see Fig. 4c). On the other hand, $\mathrm{\textit{l}}_\mathrm{2}$ norm-based FGM attack, with $\epsilon$ = 0.25, causes almost no accuracy loss in both AccL5 and M9-based AxL5 initially (see Fig. 4d). Later, the accuracy of M9-based AxL5 increases with higher values of $\epsilon$ e.g., around 0.2 but then, drops again to 65\% with $\epsilon$ = 2. This small deviating defensive behavior is exceptional and very often observed in AxDNNs \textit{due to data-dependent discontinuity of their approximation-induced errors} \cite{guesmi2020defensive}. Such discontinuity can be referred to masking and non-masking of erroneous approximation bits which can traverse through AxDNN layers.

Interestingly, Fig. \ref{fig:LenetBIMFGM}d shows a \uline{28\% accuracy loss in M9-based AxL5 but only 9\% accuracy loss in AccL5} is observed under $\mathrm{\textit{l}}_\mathrm{2}$ norm-based FGM attack with $\epsilon$ = 2. Likewise, Fig. \ref{fig:LenetPGDRAU}a shows a \uline{36\% accuracy loss in M9-based AxL5 but 17\% accuracy loss only in AccL5} is observed under the $\mathrm{\textit{l}}_\mathrm{2}$ norm-based PGD attack with $\epsilon$ = 2. Such a trend is also observed with small perturbation budgets. For example, Fig. \ref{fig:LenetPGDRAU}b illustrate that $\mathrm{\textit{l}}_\mathrm{\infty}$ norm-based PGD attack with \uline{even $\epsilon$ = 0.05 leads to 23\% accuracy loss in M9-based AxL5 but only 1\% accuracy loss in AccL5.} This identifies the non-defensive nature of AxDNNs under adversarial attacks. This observation \textit{contradicts the defensive approximation} in \cite{guesmi2020defensive}, where approximate computing was rendered defensive with such adversarial attacks. Furthermore, the BIM (see Fig. 4a and Fig. 4b) and PGD attacks (see Fig. 5a and Fig. 5b) seem to have more impact on the adversarial robustness of both AccL5 and AxL5 in comparison to the FGM attacks. In the case of AxAlx, the CIFAR-10 \cite{krizhevsky2009learning} classification under gradient-based attacks shows that their adversarial robustness is very close to AccAlx. Therefore, these results are excluded from the paper.   

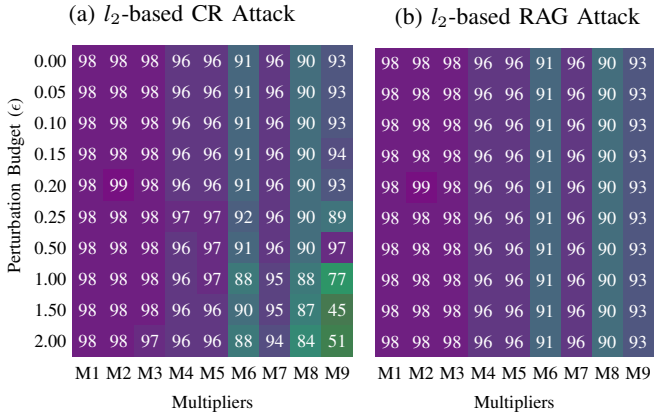
\begin{figure}[!t]
\centering
	\begin{subfigure}[b]{.26\textwidth}
		\centering
		\vspace{-0.01in}
		\caption{$\mathrm{\textit{l}}_\mathrm{2}$-based CR Attack}
		{\resizebox{\textwidth}{!}{\definecolor{mycolor2}{rgb}{0.00000,0.44700,0.74100}%
\definecolor{mycolor1}{rgb}{0.85000,0.32500,0.09800}%
\definecolor{mycolor3}{rgb}{0,128,0}%

\begin{tikzpicture}[scale=0.6]

\node at (5,-12) {Multipliers};
\node [rotate=90] at (-1.3,-5) {Perturbation Budget ($\epsilon$)};

\node at (1,-11) {M1};
\node at (2,-11) {M2};
\node at (3,-11) {M3};
\node at (4,-11) {M4};
\node at (5,-11) {M5};
\node at (6,-11) {M6};
\node at (7,-11) {M7};
\node at (8,-11) {M8};
\node at (9,-11) {M9};

\node at (0,-1) {0.00 -};
\node at (0,-2) {0.05 -};
\node at (0,-3) {0.10 -};
\node at (0,-4) {0.15 -};
\node at (0,-5) {0.20 -};
\node at (0,-6) {0.25 -};
\node at (0,-7) {0.50 -};
\node at (0,-8) {1.00 -};
\node at (0,-9) {1.50 -};
\node at (0,-10) {2.00 -};

\foreach \y [count=\n] in {
      {98,98,98,96,96,91,96,90,93},
      {98,98,98,96,96,91,96,90,93},
      {98,98,98,96,96,91,96,90,93},
      {98,98,98,96,96,91,96,90,94},
      {98,99,98,96,96,91,96,90,93},
      {98,98,98,97,97,92,96,90,89},
      {98,98,98,96,97,91,96,90,97},
      {98,98,98,96,97,88,95,88,77},
      {98,98,98,96,96,90,95,87,45},
      {98,98,97,96,96,88,94,84,51},
    } {
      \foreach \x [count=\m] in \y {
        \node[fill=violet!\x!cyan!\x!cyan!\x!cyan!\x!cyan!\x!green!\x!green!\x!black!\x!gray, minimum size=6mm, text=white] at (\m,-\n) {\x};      
      }
    }
\end{tikzpicture}}\label{fig:l2CR}}
	\end{subfigure}
	\begin{subfigure}[b]{.21\textwidth}
		\centering
		\vspace{-0.01in}
		\caption{$\mathrm{\textit{l}}_\mathrm{2}$-based RAG Attack}
		{\resizebox{\textwidth}{!}{\definecolor{mycolor2}{rgb}{0.00000,0.44700,0.74100}%
\definecolor{mycolor1}{rgb}{0.85000,0.32500,0.09800}%
\definecolor{mycolor3}{rgb}{0,128,0}%

\begin{tikzpicture}[scale=0.6]

\node at (5,-12) {Multipliers};

\node at (1,-11) {M1};
\node at (2,-11) {M2};
\node at (3,-11) {M3};
\node at (4,-11) {M4};
\node at (5,-11) {M5};
\node at (6,-11) {M6};
\node at (7,-11) {M7};
\node at (8,-11) {M8};
\node at (9,-11) {M9};

\foreach \y [count=\n] in {
      {98,98,98,96,96,91,96,90,93},
      {98,98,98,96,96,91,96,90,93},
      {98,98,98,96,96,91,96,90,93},
      {98,98,98,96,96,91,96,90,93},
      {98,99,98,96,96,91,96,90,93},
      {98,98,98,96,96,91,96,90,93},
      {98,98,98,96,96,91,96,90,93},
      {98,98,98,96,96,91,96,90,93},
      {98,98,98,96,96,91,96,90,93},
      {98,98,98,96,96,91,96,90,93},
    } {
      \foreach \x [count=\m] in \y {
        \node[fill=violet!\x!cyan!\x!cyan!\x!cyan!\x!cyan!\x!green!\x!green!\x!black!\x!gray, minimum size=6mm, text=white] at (\m,-\n) {\x};        
      }
    }
\end{tikzpicture}}\label{fig:l2RAG}}
	\end{subfigure}
	\vspace{-0.02in}
	\caption{Adversarial robustness of accurate and approximate LeNet-5 under CR and RAG attacks with the MNIST \cite{cohen2017emnist} dataset. The labels M1 to M9 refer to the 1JFF (Accurate), 96D, 12N4, 17KS, 1AGV, FTA, JQQ, L40 and JV3 multipliers in EvoApprox8b \cite{mrazek2017evoapproxsb} library.} 
	\label{fig:LenetCRRAG}
	\vspace{-0.22in}
\end{figure}

\subsubsection{\uline{Approximate DNNs under Decision-based Attacks}} 
\begin{figure*}[!t]
\centering
	\begin{subfigure}[b]{.29\textwidth}
		\centering
		\vspace{-0.01in}
		\caption{$\mathrm{\textit{l}}_\mathrm{2}$-based CR Attack}
		{\resizebox{\textwidth}{!}{\definecolor{mycolor2}{rgb}{0.00000,0.44700,0.74100}%
\definecolor{mycolor1}{rgb}{0.85000,0.32500,0.09800}%
\definecolor{mycolor3}{rgb}{0,128,0}%

\begin{tikzpicture}[scale=0.6]

\node at (5,-12) {Multipliers};
\node [rotate=90] at (-1.3,-5) {Perturbation Budget ($\epsilon$)};

\node at (1,-11) {M1};
\node at (2,-11) {M2};
\node at (3,-11) {M3};
\node at (4,-11) {M4};
\node at (5,-11) {M5};
\node at (6,-11) {M6};
\node at (7,-11) {M7};
\node at (8,-11) {M8};

\node at (0,-1) {0.00 -};
\node at (0,-2) {0.05 -};
\node at (0,-3) {0.10 -};
\node at (0,-4) {0.15 -};
\node at (0,-5) {0.20 -};
\node at (0,-6) {0.25 -};
\node at (0,-7) {0.50 -};
\node at (0,-8) {1.00 -};
\node at (0,-9) {1.50 -};
\node at (0,-10) {2.00 -};

\foreach \y [count=\n] in {
      {80,80,80,79,80,78,80,79},
      {80,80,80,79,80,78,80,79},
      {80,80,79,79,80,78,80,79},
      {80,80,78,79,80,78,80,79},
      {80,80,76,79,80,78,80,79},
      {80,80,74,79,80,78,80,78},
      {79,79,80,79,80,78,80,78},
      {77,77,80,79,79,78,79,77},
      {75,75,80,78,77,77,77,76},
      {73,73,80,76,75,76,76,75},
    } {
      \foreach \x [count=\m] in \y {
        \node[fill=violet!\x!violet!\x!violet!\x!cyan!\x!gray, minimum size=6mm, text=white] at (\m,-\n) {\x};
        }
    }
\end{tikzpicture}}\label{fig:l2CR_Alx}}
	\end{subfigure}
	\begin{subfigure}[b]{.23\textwidth}
		\centering
		\vspace{-0.01in}
		\caption{$\mathrm{\textit{l}}_\mathrm{2}$-based RAG Attack}
		{\resizebox{\textwidth}{!}{\definecolor{mycolor2}{rgb}{0.00000,0.44700,0.74100}%
\definecolor{mycolor1}{rgb}{0.85000,0.32500,0.09800}%
\definecolor{mycolor3}{rgb}{0,128,0}%

\begin{tikzpicture}[scale=0.6]

\node at (5,-12) {Multipliers};

\node at (1,-11) {M1};
\node at (2,-11) {M2};
\node at (3,-11) {M3};
\node at (4,-11) {M4};
\node at (5,-11) {M5};
\node at (6,-11) {M6};
\node at (7,-11) {M7};
\node at (8,-11) {M8};

\foreach \y [count=\n] in {
      {80,80,80,79,80,78,80,79},
      {80,80,80,79,80,78,80,79},
      {79,80,80,79,80,78,80,79},
      {79,80,80,79,80,78,80,79},
      {79,80,80,79,80,78,80,79},
      {79,80,80,79,80,78,80,79},
      {79,79,80,79,80,78,80,79},
      {79,77,78,79,79,78,79,77},
      {79,75,76,78,77,77,77,76},
      {73,73,74,76,75,76,76,75},
    } {
      \foreach \x [count=\m] in \y {
        \node[fill=violet!\x!violet!\x!violet!\x!cyan!\x!gray, minimum size=6mm, text=white] at (\m,-\n) {\x};   
      }
    }
\end{tikzpicture}}\label{fig:l2RAG_Alx}}
	\end{subfigure}
	\begin{subfigure}[b]{.23\textwidth}
		\centering
		\vspace{-0.01in}
		\caption{$\mathrm{\textit{l}}_\mathrm{2}$-based RAU Attack}
		{\resizebox{\textwidth}{!}{\definecolor{mycolor2}{rgb}{0.00000,0.44700,0.74100}%
\definecolor{mycolor1}{rgb}{0.85000,0.32500,0.09800}%
\definecolor{mycolor3}{rgb}{0,128,0}%

\begin{tikzpicture}[scale=0.6]

\node at (5,-12) {Multipliers};

\node at (1,-11) {M1};
\node at (2,-11) {M2};
\node at (3,-11) {M3};
\node at (4,-11) {M4};
\node at (5,-11) {M5};
\node at (6,-11) {M6};
\node at (7,-11) {M7};
\node at (8,-11) {M8};

\foreach \y [count=\n] in {
      {80,80,80,79,80,78,78,79},
      {80,80,80,79,80,78,78,79},
      {80,80,80,79,80,78,78,79},
      {80,80,80,79,80,78,78,79},
      {80,80,80,79,80,78,78,79},
      {80,80,80,79,80,78,78,78},
      {79,79,80,79,80,78,78,78},
      {77,77,78,79,79,77,77,78},
      {75,75,76,78,78,77,77,77},
      {73,73,74,76,76,76,75,75},
    } {
      \foreach \x [count=\m] in \y {
        \node[fill=violet!\x!violet!\x!violet!\x!cyan!\x!gray, minimum size=6mm, text=white] at (\m,-\n) {\x};  
      }
    }
\end{tikzpicture}}\label{fig:l2RAU_Alx}}
	\end{subfigure}
	\begin{subfigure}[b]{.23\textwidth}
		\centering
		\vspace{-0.01in}		\caption{$\mathrm{\textit{l}}_\mathrm{\infty}$-based RAU Attack}
		{\resizebox{\textwidth}{!}{\definecolor{mycolor2}{rgb}{0.00000,0.44700,0.74100}%
\definecolor{mycolor1}{rgb}{0.85000,0.32500,0.09800}%
\definecolor{mycolor3}{rgb}{0,128,0}%

\begin{tikzpicture}[scale=0.6]

\node at (5,-12) {Multipliers};

\node at (1,-11) {M1};
\node at (2,-11) {M2};
\node at (3,-11) {M3};
\node at (4,-11) {M4};
\node at (5,-11) {M5};
\node at (6,-11) {M6};
\node at (7,-11) {M7};
\node at (8,-11) {M8};

\foreach \y [count=\n] in {
      {80,80,80,79,80,78,80,79},
      {74,74,75,77,76,76,77,76},
      {67,67,68,72,70,73,70,71},
      {57,58,59,64,62,66,62,64},
      {47,47,49,55,52,58,54,56},
      {37,37,40,47,43,50,43,43},
      {8,8,10,17,12,22,13,24},
      {0,0,0,0,0,0,0,0},
      {0,0,0,0,0,0,0,0},
      {0,0,0,0,0,0,0,0},
    } {
      \foreach \x [count=\m] in \y {
        \node[fill=violet!\x!violet!\x!violet!\x!cyan!\x!gray, minimum size=6mm, text=white] at (\m,-\n) {\x}; 
      }
    }
\end{tikzpicture}}\label{fig:linfRAU_Alx}}
	\end{subfigure}
	\vspace{-0.24in}
	\caption{Adversarial robustness of accurate and approximate Alexnet under CR, RAG and RAU attacks with the CIFAR-10 \cite{krizhevsky2009learning} dataset. The labels M1 to M9 refer to the 1JFF (Accurate), 2P7, KEM, 150Q, 14VP, QJD, 1446 and GS2 multipliers in EvoApprox8b \cite{mrazek2017evoapproxsb} library.} 
	\label{fig:alexnet}
	\vspace{-0.1in}
\end{figure*}
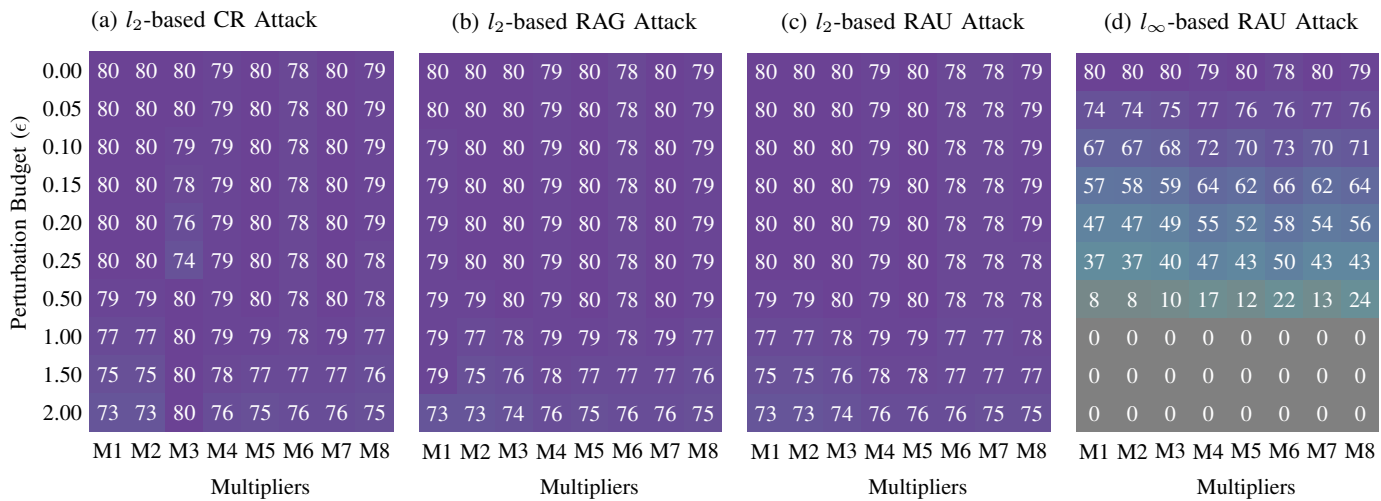

The decision-based attacks affect the adversarial robustness of both AxDNNs and accurate DNNs. For example, Fig. \ref{fig:LenetPGDRAU}d shows that $\mathrm{\textit{l}}_\mathrm{\infty}$ norm-based RAU attack, with $\epsilon$ = 1, leads to 50\% accuracy loss in AccL5. However, the same attack leads to 75\% accuracy loss in M8-based AxL5. The same trend is observed in the case of other decision-based attacks. It is also noticed that $\mathrm{\textit{l}}_\mathrm{2}$ norm-based decision-based attacks are comparatively less perturbing. However, their impact on the accuracy of AxDNNs is higher with an increase in the approximation noise in AxDNNs. Interestingly, $\mathrm{\textit{l}}_\mathrm{2}$-based CR attack in Fig. 6a undergoes \uline{almost no accuracy loss (as low as 0.06\%) in AccL5} in spite of high perturbation budget i.e., $\epsilon$ = 1.5. The same attack has \uline{53\% accuracy loss in M8-based AxL5}. Small deviations in these trends are observed with decision-based attacks as well but they do not cause significant changes in the accuracy. Moreover, a similar adversarial robustness trend is observed with CIFAR-10 \cite{krizhevsky2009learning} classification in the case of decision-based attacks on Alexnet as shown in Fig. \ref{fig:alexnet}. AxAlx performs close to AccAlx but the impact of decision-based attacks is more noticeable in $\mathrm{\textit{l}}_\mathrm{\infty}$ norm-based RAU attack (see Fig. 7d).  

\subsection{Transferability Analysis} 
\label{subsec:transfer}
Adversarial examples are known for their transferability~\cite{adrel7}, which means that they are transferable from one model to another model. This also refers to the fact that it is possible to attack models to which the attacker does not have access~\cite{demontis2019adversarial}. Therefore, for further analyzing the adversarial robustness of AxDNNs, we craft the adversarial examples using accurate DNN models (second attack scenario discussed in Section \ref{subsec:advknow}) and evaluate their impact on AxDNNs with different model structures. Our results show that the adversarial attacks are more transferable if the adversary is not aware of both the inexactness and type of DNN model used in the inference engine. For example, Table \ref{tab:transfer} shows that $\mathrm{\textit{l}}_\mathrm{\infty}$ norm-based BIM attack ($\epsilon$ = 0.05), strongest attack in previous section, is more transferable from AccL5 to AxAlx when compared to AxL5, and AccAlx to AxL5 when compared to AxAlx. The same trend is observed with both MNIST \cite{cohen2017emnist} and CIFAR-10 \cite{krizhevsky2009learning} datasets. 

\begin{figure}[!t]
	\centering
	\vspace{-0.05in}
	\includegraphics[width=1\linewidth]{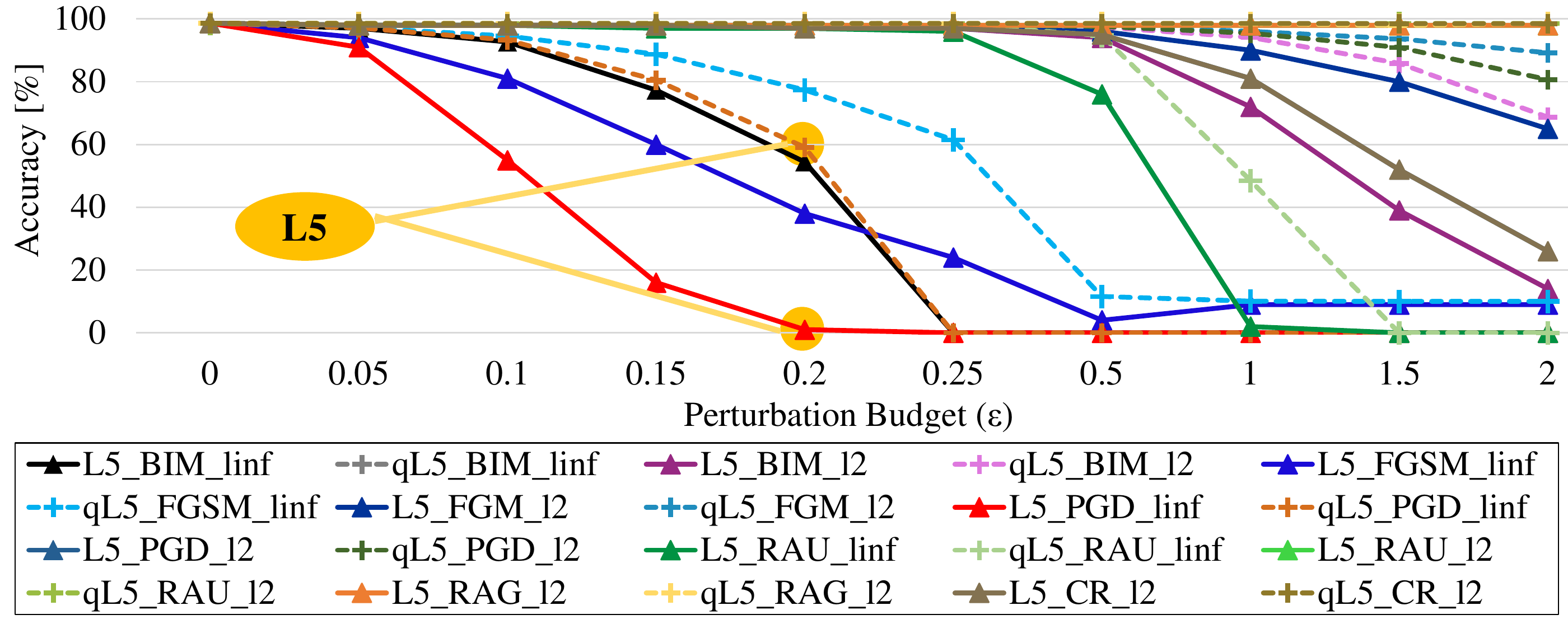}
	\caption{Adversarial robustness of quantized (qL5) and non-quantized accurate Lenet-5 (L5) using the MNIST \cite{cohen2017emnist}}
	\label{fig:qAcc}
	\vspace{-0.15in}
\end{figure}

\begin{table}[!t]
\vspace{0.01in}
\caption {Transferability Analysis with $\mathrm{\textit{l}}_\mathrm{\infty}$ norm-based BIM attack ($\epsilon$ = 0.05). X/Y represents accuracy before/after attack}
\label{tab:transfer}
\vspace{0.1in}
\centering
\begin{tabular}{|c|cc|cc|}
\hline
\rowcolor[HTML]{E0E0E0} 
\cellcolor[HTML]{E0E0E0}{\color[HTML]{000000} }                                                                        & \multicolumn{2}{c|}{\cellcolor[HTML]{E0E0E0}{MNIST \cite{cohen2017emnist}}}                               & \multicolumn{2}{c|}{\cellcolor[HTML]{E0E0E0}{CIFAR-10 \cite{krizhevsky2009learning}}}                               \\ \cline{2-5} 
 
\multirow{-2}{*}{\cellcolor[HTML]{E0E0E0}{\begin{tabular}[c]{@{}c@{}}DNN \\ Models\end{tabular}}} & \multicolumn{1}{c|}{{AxL5}} & {AxAlx} & \multicolumn{1}{c|}{{AxL5}} & {AxAlx} \\ \hline
~AccL5~                                                                                                                  & \multicolumn{1}{c|}{~98/97~}                                               & ~67/43~                        & \multicolumn{1}{c|}{~54/9~}                                                & ~53/4~                         \\ \hline
~AxAlx~                                                                                                                  & \multicolumn{1}{c|}{~98/9~}                                                & ~67/11~                        & \multicolumn{1}{c|}{~54/20~}                                               & ~53/10~                        \\ \hline
\end{tabular}
\vspace{-0.15in}
\end{table}

\subsection{Adversarial Quantization and Approximation} 
\label{subsec:advquan}
Fig. \ref{fig:LenetBIMFGM} - Fig. \ref{fig:alexnet} present AxDNNs which employ approximate computing along-with quantization. From their comparison with 8-bit quantized accurate DNNs in Fig. \ref{fig:qAcc}, we observe that quantization improves the adversarial robustness \cite{khalid2019qusecnets}. However, approximate computing does not support this behavior. The classification accuracy of AxDNNs decreases with an increase in the strength of the adversarial attacks in spite of employing quantization. For example, under $\mathrm{\textit{l}}_\mathrm{\infty}$ norm-based PGD attack ($\epsilon$ = 0.2), the quantization increases the accuracy of non-quantized AccL5 by 58\% in Fig. \ref{fig:qAcc} (see label L5). Conversely, approximate computing decreases the accuracy of quantized AccL5 by 35\% in M8-based AxL5 under the same attack in Fig. \ref{fig:LenetPGDRAU}b. The self-error-inducing nature of approximate computing degrades the performance of quantized DNN models and hence, leads to successful adversarial attacks. Thus, approximate computing acts antagonistically to quantization under the adversarial attacks. \\

\noindent \textit{Summary.} Most state-of-the-art AxDNNs employ approximate multipliers for reducing their energy consumption \cite{riaz2020caxcnn} \cite{marchisio2020red} \cite{spantidi2021positive}. These approximate multipliers either have approximate partial products generation or addition. However, approximation error in both cases depends on the \textit{specific} input bit combinations \cite{mazahir2019probabilistic}. Recent work of Gusemi et al. \cite{guesmi2020defensive} exploited such approximation behavior in AxDNN for the defense against the fixed strength of the adversarial attacks. However, such defensive nature of AxDNNs is not always observable as shown by our experimental results in Section \ref{sec:results}. In a real-world scenario, the adversary can vary the perturbation budget which may lead to successful misclassification. 
Interestingly, AxDNNs are not only vulnerable to higher perturbation budgets but also to very small perturbations budgets in some cases, such as $\epsilon$ =0.05 as shown in our experimental results. Note, an attack with such a small perturbations budget can be stealthy enough to bypass the attack detection techniques and remain imperceptible to the human eye.



\section{Conclusion}
\label{sec:conclusion}

In this paper, we explored the adversarial robustness of AxDNNs, using the state-of-the-art unsigned approximate multipliers, with the MNIST and CIFAR10 datasets. We empirically show that an adversarial attack on AxDNN leads to 53\% accuracy loss. Conversely, the same attack leads to almost no accuracy loss (as low as 0.06\%) in accurate DNNs. We observe that approximate computing can reduce the adversarial robustness in spite of quantization in AxDNNs and partial knowledge of the adversary about the model structure. In summary, this work answers three research questions in Section \ref{sec:introduction} as follows:\\

\noindent \textbf{(A1)} Though AxDNNs often exhibit mere defensive behavior; this trend is not universal (or consistent). Their adversarial robustness decreases with an increase in the perturbation budget and occasionally, surpasses the accurate DNNs. They are \textit{not universally defensive} in nature towards the adversarial attacks. \\
\noindent \textbf{(A2)} The adversarial attacks are \textit{transferable from accurate DNNs to AxDNNs} even if the adversary has partial knowledge about their inexactness and model structure. \\
\noindent \textbf{(A3)} The quantization and approximate computing act \textit{antagonistic} to each other in  an adversarial environment.


\bibliographystyle{IEEEtran}
\bibliography{bib/conf}

\end{document}